\documentclass[12pt]{article}

\newtheorem{one}{Definition}[section]{\bf}{\em}
\newtheorem{three}[one]{Definition}{\bf}{\em}
\newtheorem{three1}{Lemma}[section]{\bf}{\em}
\newtheorem{three3}[three1]{Corollary}{\bf}{\em}
\newtheorem{three2}[three1]{Corollary}{\bf}{\em}
\newtheorem{seven}[one]{Definition}{\bf}{\em}
\newtheorem{prop1}{Proposition}[section]{\bf}{\em}
\newtheorem{prop2}{Proposition}[section]{\bf}{\em}

\begin{document}

\title{Stationarity-Conservation Laws for
Certain Linear Fractional
Differential Equations}
\author{Ma\l gorzata Klimek \\
Institute of Mathematics and Computer Science \\
Technical University of Cz\c{e}stochowa, \\
ul D\c{a}browskiego 73, 42-200
Cz\c{e}stochowa, Poland \\
e-mail: klimek@matinf.pcz.czest.pl}
\date{}
\maketitle
\newpage
\abstract{The Leibniz's rule for fractional Riemann-Liouville derivative is studied
in algebra of functions defined by Laplace convolution. This algebra and the derived
 Leibniz's rule is used in construction of explicit form of
stationary-conserved currents for linear fractional differential equations. The examples of
fractional diffusion in 1+1 and the fractional diffusion in d+1 dimensions are discussed in detail.
The results are generalized to the mixed fractional-differential
and mixed sequential fractional-differential systems for which the stationarity-conservation laws
are obtained. The derived currents are used in construction of stationary nonlocal
charges.       
}
\newpage
\section{Introduction}
In the paper we shall study the properties of the fractional-differential equations
together with the mixed models containing both fractional and standard classical derivatives. \\
The fractional analysis describing the fractional integrals and derivatives is covered extensively in literature (see
for example the monographies \cite{OS,MR,PD,SK} an references given therein). Recently
these
operators have found application in various areas of physics.
Let us start with fractional mechanics describing the nonconservative systems developed
by Riewe \cite{R1,R2} who also shows the possible connection
between the fractional formalism and a problem of classical frictional force proportional to velocity. \\
The fractional operators emerge also  as the infinitesimal generators
of coarse grained macroscopic time evolutions \cite{H1,H2,H3,H4,HS}
and determine fractional diffusion processes \cite{SK,NG,WW,WS,CO,M}.\\
The phenomenological approach to derivation of the stress-strain relationships
which tends to proper description of the rheological properties
of wide classes of materials leads to rheological constitutive equations
with fractional derivatives \cite{SF}. \\
Next domain is the path-integral formulation of classical boundary problems with fractal boundaries
used in polymer science. These models can be rewritten in form of fractional differential equations.
The order of the fractional operator is given by the geometry of the boundary, the space in which the boundaries
are embedded and the type of random walk process \cite{SF,DG}. \\
One should also mention the description of wandering processes given by the fractional Fokker-Planck-Kolmogorov
equation in the fractal space-time \cite{F1,Z1,Z2,F2},
the fractional generalization of Klein-Kramers equation which yields the fractional Raleigh and
Fokker-Planck models \cite{M1,M2,M3} and
the fractional equation describing the end-to-end distribution of stable random walk
where the fractional power of the standard Laplace operator is used \cite{DG}. \\
Finally the fractional operators appear also in field theory where recently
the roots of the wave operator were investigated by Zavada \cite{ZP,Z}.
In his paper he shows that the Dirac operator is the only one from them which can
be realized using the standard derivatives. When the root of order different from
$ \frac{1}{2}$ is considered we obtain the fractional differential equation. \\
Most of the above examples are linear equations with constant coefficients of mixed type - containing
both fractional and standard derivatives. As is well-known in classical field theory the conservation laws
for linear differential systems can be derived using Takahashi-Umezawa method \cite{TU}. This procedure
has been extended to discrete and noncommutative models \cite{k1,k2,k3}. \\
Our aim is to show that similar procedure can be applied to fractional equations
in the convolution algebra of functions in order to obtain
the stationarity-conservation laws which are analogs of conservation
equations known for models from classical differential calculus.
The  explicitly derived  stationary-conserved currents are nonlocal expressions
with respect to this part of space for which the fractional derivatives appear in the
initial equation. This phenomenon is connected
 with the nonlocality of fractional operators as well as with the convolution
algebra of functions which we introduce to
simplify the Leibniz's rule in the fractional differential calculus. \\
Some of the derived nonlocal currents yield the stationary charges which in turn
can be converted into   nonlocal conserved charges.
In the present paper we discuss this procedure on some examples and then for general case
of mixed fractional and differential equations.
These nonlocal integro-differential equations
obey new type of conservation law which we call stationarity-conservation law. \\
In section 2 we review briefly the properties of Riemann-Liouville fractional integrals and derivatives
and show that the Leibniz's rule
is simplified in the algebra of convolution. The new Leibniz's rule produces
strict requirements concerning the behaviour of the functions
in the neighbourhood of 0. The next section  contains the detailed discussion of the derivation the stationarity-conservation law
for  two examples of fractional
diffusion: in 1+1 and d+1 dimensions.
It is explicitly proven that the asymptotic properties of the solutions for
diffusion equation in 1+1 dimension
allow construction of stationary currents and stationary charges.
 Then the currents and charges are converted via convolution to conserved currents
and charges which are stationary in a strict sense - that means true constant functions. \\
Final section includes the general fractional-differential model as well as
the sequential fractional-differential one.
We show explicit construction of stationary currents, the derivation of the stationarity-conservation
laws and close the section with discussion of the possible stationary and conserved charges.
\section{Properties of fractional integrals and derivatives}
\subsection{Riemann-Liouville fractional integral}
Let us recall the definition of Riemann-Liouville fractional integral \cite{OS,MR,PD}
used widely in the literature dealing with fractional calculus:
\begin{one}
Let $ Re \nu>0 $ and let $ f $ be piecewise continuous on $ (0, + \infty ) $
and integrable on any finite subinterval of $ [0, + \infty) $. Then for $ t>0 $
\begin{equation}
D^{-\nu}_{t} f(t): =\frac{1}{\Gamma(\nu)} \int_{0}^{t} (t-s)^{\nu-1}f(s) ds
\label{int}
\end{equation}
is the Riemann-Liouville fractional integral of $ f $ of order $ \nu $.
\end{one}
We notice that the above definition includes the operation of
Laplace convolution, namely it can be written as:
\begin{equation}
 D^{-\nu}_{t} f(t)=\Phi_{-\nu} * f (t)=f* \Phi_{-\nu}(t)
 \label{iconv}
\end{equation}
where we have denoted $ \Phi_{-\nu}(t)=\frac{1}{\Gamma(\nu)} t^{\nu-1} $. \\
Now we are interested in the properties of the fractional integral connected with the composition
of the integrals with respect to the same coordinate. The answer is the generalization
of the Dirichlet's integral formula for continuous function which is called in fractional calculus
the composition rule \cite{OS,MR,PD}:
\begin{equation}
D^{-\nu}D^{-\mu} f(t)=D^{-(\nu+\mu)}f(t)=D^{-\mu}D^{-\nu} f(t)
\label{icomp}
\end{equation}
for $ Re\mu, Re\nu >0 $ and for any function $ f $ piecewise continuous on $ [0, +\infty) $.\\
Let us now present the known forms of Leibniz's rule for integral (\ref{int}):
\begin{equation}
D^{-\nu} (f \cdot g)=\sum_{j=0}^{\infty} \left(\stackrel{-\nu}{j} \right ) D^{-\nu-j}f \cdot D^{j}g
\label{leib1}
\end{equation}
where $ f $ and $ g $ are real analytic functions on $ [0, +\infty) $. This rule was generalized
by Osler \cite{OS,O1,O2,O3} who obtained
the following forms of Leibniz's rule:
\begin{equation}
D^{-\nu} (f \cdot g)=\sum_{j=-\infty}^{+\infty} \frac{\Gamma(-\nu+1)}{\Gamma(-\nu-\gamma-j+1)\Gamma(\gamma +j+1)}
 D^{-\nu-\gamma-j}f\cdot D^{\gamma+j}g
 \label{leib2}
\end{equation}
\begin{equation}
D^{-\nu} (f \cdot g)=\int_{-\infty}^{+\infty}\frac{\Gamma(-\nu+1)}
{\Gamma(-\nu-\gamma-\lambda+1)\Gamma(\gamma +\lambda+1)}
 D^{-\nu-\gamma-\lambda}f\cdot D^{\gamma+\lambda}g d\lambda
 \label{leib3}
\end{equation}
where $ \gamma $ is an arbitrary complex number. \\
We shall not discuss the convergence of the series in (\ref{leib2})
 and of the improper integral in  (\ref{leib3}). Let us notice however
that when the algebra of functions is defined by standard point-wise multiplication as in the above formulas
all versions of Leibniz's rule are very complicated. \\
Thus we propose to investigate the algebra of functions with multiplication defined via Laplace convolution:
\begin{equation}
f*g (t):=\int_{0}^{t} f(t-s)g(s)ds
\label{mult}
\end{equation}
As is well known this multiplication is associative and commutative . The neutral element
is the Dirac $ \delta$-function. Let us prove the following Leibniz's rule for fractional integral (\ref{int})
and multiplication defined by (\ref{mult}):
\begin{equation}
D^{-\nu} (f*g)=(D^{-(\nu-\gamma)}f)*D^{-\gamma}g
\label{trans}
\end{equation}
where $ Re \nu>0 $ and $ \gamma $ a complex number fulfilling inequality $Re(\nu-\gamma)\geq 0 $.\\
The new Leibniz's rule is implied by the composition rule (\ref{icomp}) and properties of convolution
which defines the fractional integral (\ref{iconv}) and algebra of functions (\ref{mult}):
\begin{equation}
D^{-\nu} (f*g)=D^{-\gamma-(\nu-\gamma)}(f*g)=D^{-\gamma}D^{-(\nu-\gamma)}( f*g)=
\end{equation}
$$
\left (\left ((f*g)*\Phi_{-(\nu-\gamma)}\right )*\Phi_{-\gamma}\right )=
(f* \Phi_{-(\nu-\gamma)}) *(g*\Phi_{-\gamma})=
$$
$$
=D^{-(\nu-\gamma)}f*D^{-\gamma}g
$$
where $ Re \nu>0 $ and $Re(\nu-\gamma)\geq 0 $. \\
The derived formula (\ref{trans}) is similar to the multiplicity properties of the transformation operators
in the discrete \cite{k1} and noncommutative \cite{k2,k3} differential multidimensional calculi for standard product
of functions:
\begin{equation}
\zeta^{i}_{j} (f \cdot g)=(\zeta^{i}_{k}f)\cdot (\zeta^{k}_{j}g)
\label{dis}
\end{equation}
The multiplicity property of the fractional calculus (\ref{trans}) leads to the following redefinition of the integral of order $ \nu $:
\begin{equation}
{\cal{D}}^{-\nu}_{t} f(t):=(D^{-\nu}_{t}-1)f(t)=f*(\Phi_{-\nu}-\delta)(t)
\label{opn}
\end{equation}
The new operator $ \cal{D} $ obeys the following Leibniz's rule in the algebra
defined by convolution product (\ref{mult}):
\begin{equation}
{\cal{D}}^{-\nu}_{t}(f*g)=(\zeta^{-\gamma}f)*{\cal{D}}^{-(\nu-\gamma)}_{t} g+({\cal{D}}^{-\gamma}_{t}f)*g
\label{ileib1}
\end{equation}
or its symmetric form:
\begin{equation}
{\cal{D}}^{-\nu}_{t}(f*g)=f*{\cal{D}}^{-\gamma}_{t} g+({\cal{D}}^{-(\nu-\gamma)}_{t}f)*\zeta^{-\gamma}g
\label{ileib2}
\end{equation}
where for given $ \nu $ the $ \gamma $ is a complex number fulfilling conditions:
$ Re \gamma>0 $, $ Re(\nu-\gamma) \geq 0 $. As we have
noticed the analogy between the action of the fractional operator $ D $ in the algebra
of convolution product (\ref{mult}) and the transformation operator $ \zeta $
in the dicrete and noncommutative algebra (\ref{dis}) we shall use in the sequel the notation $ \zeta $ for the "old"  fractional integral
(\ref{int}):
\begin{equation}
\zeta^{-\alpha}\equiv D^{-\alpha}_{t}
\label{itrans}
\end{equation}
The above Leibniz's rules are implied by the properties of the convolution and the composition rule
(\ref{icomp}). If $ \nu $ and $ \gamma $ fulfill the above restrictions we obtain:
\begin{equation}
{\cal{D}}^{-\nu}_{t} (f*g)=
\end{equation}
$$ f*g*(\Phi_{-\nu} -\delta)=f*g*(\Phi_{-\gamma-(\nu-\gamma)}\pm \Phi_{-\gamma}-\delta) = $$
$$ f*g*\Phi_{-\gamma} *(\Phi_{-(\nu-\gamma)}-\delta)+f*g*(\Phi_{-\gamma}-\delta)= $$
$$ (f*\Phi_{-\gamma})*g*(\Phi_{-(\nu-\gamma)}-\delta)+ f*(\Phi_{-\gamma}-\delta)*g= $$
$$ (\zeta^{-\gamma}f)*{\cal{D}}^{-(\nu-\gamma)}_{t} g+({\cal{D}}^{-\gamma}_{t}f)*g $$
The proof of the symmetric form of the Leibniz's rule (\ref{ileib2}) is analogous.
\subsection{Riemann-Liouville fractional derivative}
The operator known as the Riemann-Liouville fractional derivative \cite{OS,MR,PD}
is defined using the fractional integral (\ref{int}):
\begin{three}
Let $ m \leq Re \nu < m+1 $, $ t>0$. The operator given by formula:
\begin{equation}
D^{\nu}_{t}:=\left (\frac{d}{dt}\right )^{m+1}D^{-(m-\nu+1)}_{t}f(t)
\label{der}
\end{equation}
for functions for which the improper integral on the right-hand side of (\ref{der}) is convergent
is  called the Riemann-Liouville fractional derivative of order $ \nu $.
\end{three}
Let us notice that the functions from the domain of the $ D ^{\nu}_{t} $ operator
form the subset in the set of functions from definition 2.1. It is well-known fact that
this class consists of
finite sums of functions of the type:
\begin{equation}
t^{\lambda}\sum_{k=0}^{\infty} a_{k}t^{k}
\label{type1}
\end{equation}
or
\begin{equation}
t^{\lambda}ln(t) \sum_{k=0}^{\infty} a_{k}t^{k}
\label{type2}
\end{equation}
where $ Re \lambda>-1 $ and the series have a positive radius of convergence.\\
Contrary to the fractional integrals the derivative (\ref{der}) cannot be expressed
using only convolution. The formula includes the classical derivative
 and looks as follows:
\begin{equation}
D^{\nu}_{t} f(t):= \left (\frac{d}{dt}\right )^{m+1} (f*\Phi_{\nu-m} (t))
\label{iconvd}
\end{equation}
with the function $ \Phi_{\nu-m}=\frac{t^{-\nu+m}}{\Gamma(m+1-\nu)} $. \\
We expect the fractional derivative to obey the composition rule analogous
to the one for fractional integral. In fact \cite{OS,MR} the following formula which
generalizes (\ref{icomp}) is valid:
\begin{equation}
D^{\nu}_{t}D^{\mu}_{t} f=D^{\nu+\mu}_{t} f
\label{icompd}
\end{equation}
provided:
\begin{itemize}
\item{$ \nu$ arbitrary, $ \mu<\lambda+1 $ and the function $ f$ is of the type described by (\ref{type1},\ref{type2})}
\item{$ \nu$ arbitrary, $ \mu \ge \lambda+1 $ and $ a_{k}=0 \;\;\;\; k=0,...m-1 $ for the function $ f $
of type (\ref{type1},\ref{type2}) where $ m $ is the smallest integer greater or equal to $ Re \mu $.}
\end{itemize}
The above formula shows that the fractional derivatives
of different orders do not always commute as it is the case with
the fractional integrals. \\
The Leibniz's rule for fractional derivative has the form ($Re \nu \leq n-1 $) \cite{OS,MR,O1,O2,O3}:
\begin{equation}
D^{\nu}_{t} f \cdot g (t)=\sum_{k=0}^{n} \left (\stackrel{\nu}{k}\right ) g^{(k)} \cdot D^{\nu -k}_{t}f (t) -R^{\nu}_{n}(t)
\label{dleib}
\end{equation}
when the  function $ f $ is continuous
in the interval $ [0,t] $ while $ g $ has $ n+1 $ continuous derivatives in $ [0,t]$. \\
The remainder $ R _{n} $ is the integral expression:
\begin{equation}
R^{\nu}_{n}(t)=\frac{1}{n! \Gamma(-\nu)}\int_{0}^{t} (t-s)^{\nu-1}f(s)ds \int_{s}^{t}g^{(n+1)}(\omega) (s-\omega)^{n} d\omega
\end{equation}
If the above remainder goes to $ 0 $ for $ n \rightarrow \infty $ the Leibniz's rule (\ref{dleib})
can be written for analytic functions in the form of series:
\begin{equation}
 D^{\nu}_{t} f \cdot g=\sum_{k=0}^{\infty} \left (\stackrel{\nu}{k}\right ) f^{(k)} \cdot D^{\nu -k}_{t}g
\end{equation}
Again the form of Leibniz's rules for the algebra
defined by point-wise multiplication of functions is complicated. \\
We propose to use the algebra of convolution (\ref{mult}). The following
statment is valid for the new algebra of functions:
\begin{three1}
Let $ m \leq Re \nu < m+1 $ and the function $ g $
be piecewise continuous in $ (0, +\infty)$. If the function $ f $ is a finite sum of functions of the type (\ref{type1}, \ref{type2})
and fulfills the condition:
$$ \lim_{t \rightarrow 0+0} f^{(k)}*\Phi_{\nu-m}=0 $$
for $ k=0,1..,m $
then the following rule holds:
\begin{equation}
D^{\nu}_{t}(f * g)=(D^{\nu}_{t}f)*g
\label{formula}
\end{equation}
\end{three1}
Proof: \\
We use the well-known rule for differentiation of an integral depending on a parameter
with the upper limit depending on the same parameter:
\begin{equation}
\frac{d}{dt} \int_{0}^{t} F(t,s)ds =\int_{0}^{t}\frac{\partial F(t,s)}{\partial t} ds +
\lim_{s \rightarrow t-0} F(t,s)
\label{rule}
\end{equation}
and from it follows for $ 0 < Re \nu <1 $:
\begin{equation}
D^{\nu}_{t}(f * g)=\frac{d}{dt} (f*g*\Phi_{\nu})=\frac{d}{dt} (f*\Phi_{\nu}*g)=\left (\frac{d}{dt} (f*\Phi_{\nu})\right )*g=(D^{\nu}_{t}f)*g
\end{equation}
provided:
\begin{equation}
\lim_{t \rightarrow 0+0 } f*\Phi_{\nu}(t) =0
\end{equation}
Thus when the assumptions are fulfilled the formula (\ref{formula}) is valid. \\
For $ m <  Re \nu <m+1 $ we apply the rule (\ref{rule}) $ m+1 $ times:
$$ D^{\nu}_{t}(f*g)=\left (\frac{d}{dt}\right )^{m+1}(f*g*\Phi_{\nu-m}) =
 \left (\frac{d}{dt}\right )^{m+1}[(f*\Phi_{\nu-m})*g] =
$$
$$
 \left (\frac{d}{dt}\right )^{m}\left [[\frac{d}{dt}(f*\Phi_{\nu-m})]*g \right] = ...=
 \left [ \left (\frac{d}{dt}\right )^{m+1} (f*\Phi_{\nu-m})\right ]*g \
$$
  and arrive at the
conditions:
\begin{eqnarray}
& \lim_{t \rightarrow 0+0 } f*\Phi_{\nu-m}(t) =0 & \label{f1} \\
& \lim_{t \rightarrow 0+0 } f'*\Phi_{\nu-m}(t) =0 & \label{f2} \\
& ... & \\
& \lim_{t \rightarrow 0+0 } f^{(m)}*\Phi_{\nu-m}(t) =0 & \label{f}
\end{eqnarray}
which are fulfilled by assumption. \\
The above set of right-sided limits determines the behaviour
of the function $ f $ in the neighbourhood of $ t=0 $, namely
$ f(t) \sim t^{\beta} $ with $ \beta $ a complex number fulfilling
the condition: $ Re \beta>-1+Re \nu $ .\\
\\
The symmetric version of the formula (\ref{formula}) follows from the commutativity of the Laplace convolution.
\begin{three3}
Let $ m \leq Re \nu < m+1 $ and functions
$ f $ and $ g $ are piecewise continuous in $ (0, +\infty) $. If both functions $ f,g $  are finite sums of functions of the type (\ref{type1}, \ref{type2})
and both of them obey the assumptions from Lemma 2.1
then the following rule holds:
\begin{equation}
D^{\nu}_{t}(f * g)=\beta (D^{\nu}_{t}f)*g+ (1-\beta)  f*(D^{\nu}_{t}g)
\label{formula2}
\end{equation}
for $ \beta \in [0,1]$
\end{three3}
The above lemma together with the composition rule (\ref{icompd})
yields the analog of the property (\ref{trans}) for  Riemann-Liouville
fractional derivative:
\begin{three2}
Let  $ Re \nu >0$ and
the function $ f*g $ obey for certain $ \gamma $,
fulfilling $ Re \gamma >0 $ and $ Re (\nu-\gamma) >0 $, the assumptions
of the composition rule (\ref{icompd}). If function $ f $ fulfills the conditions from Lemma 2.1
for $ \gamma $ and the function $ g$ the corresponding  conditions for
$ \nu-\gamma$ then the following formula holds:
\begin{equation}
D^{\nu}_{t}( f*g)=D^{\nu-\gamma}_{t}D^{\gamma}_{t}(f*g)=(D^{\gamma}_{t}f)*D^{\nu-\gamma}_{t}g
\end{equation}
\end{three2}
Analogously to (\ref{opn}) we can introduce the new differintegral operator:
\begin{equation}
{\cal{D}}^{\nu}_{t}f(t):=(D^{\nu}_{t}-1)f(t)
\label{opp}
\end{equation}
The Leibniz's rule for the introduced differintegrable operator
of positive order $ \nu $ is similar to the one known from the discrete and noncommutative
calculus \cite{k1,k2,k3}:
\begin{eqnarray}
& {\cal{D}}^{\nu}_{t}(f*g)=({\cal{D}}^{\gamma}_{t}f)*g+(\zeta^{\gamma}f)*{\cal{D}}^{\nu-\gamma}_{t}g& \label{dleib1}\\
& {\cal{D}}^{\nu}_{t}(f*g)=({\cal{D}}^{\gamma}_{t}f)*\zeta^{\nu-\gamma}g+f*{\cal{D}}^{\nu-\gamma}_{t}g &  \label{dleib2}
\end{eqnarray}
where we use the notation:
\begin{equation}
\zeta^{\gamma}\equiv D^{\gamma}_{t}
\label{dtrans}
\end{equation}
and $ \nu $, $ \gamma$ together with functions $ f,g$  fulfill the conditions from Lemma 2.1.
\subsection{Riemann-Liouville partial fractional derivatives}
Let us extend the formalism introduced in previuos sections to multidimensional case. We shall study
the stationarity-conservation equations for some fractional partial differential equations and
derive for them the explicit form of stationary currents
connected with their symmetries. We assume that in the equation both types of derivatives can appear - fractional with respect to
to a subset of coordinates and classical - continuous ones with respect to the rest of coordinates.\\
Thus the question arises how to define the multiplication of functions. We propose
to use the multidimensional Laplace convolution when initial  equation contains only
Riemann-Liouville fractional partial derivatives of the form:
\begin{equation}
D^{\alpha_{k}}_{k} f(\vec{x}):=
\label{dpar}
\end{equation}
$$
\frac{1}{\Gamma(m_{k}+1-\alpha_{k})}
\left (\partial_{x_{k}} \right )^{m_{k}+1} \int_{0}^{x_{k}}
(x_{k}-s)^{-\alpha_{k}+m_{k}} f(\vec{x}+(s-x_{k})\vec{e}_{k}) ds
$$
where $ m_{k} \leq Re \alpha_{k}<m_{k}+1 $.
The upper index in the formula denotes the fractional order of the partial derivative
 while the lower one  says that it was taken with respect to coordinate $ x_{k}$. \\
 Let $ x_{1},...,x_{m} $ be a subset of coordinates in our n-dimensional model for which the fractional partial
 derivatives (\ref{dpar}) appear in the equation. Then we define multiplication of functions as follows:
 \begin{seven}
 The algebra of functions is defined by the multiplication formula:
 \begin{equation}
f*g(\vec{x}):=
\label{alg}
\end{equation}
$$
\int_{0}^{x_{1}}... \int_{0}^{x_{m}} f\left (\vec{x}-\sum_{l=1}^{m}s_{l}\vec{e}_{l}\right)
g\left (\vec{x}+\sum_{l=1}^{m}(s_{l}-x_{l})\vec{e}_{l} \right) ds_{1}...ds_{m}
$$
where $ (\vec{e}_{l})_{k}=\delta_{lk} $.
 \end{seven}
 Similarly to the one-dimensional case the multiplication (\ref{alg})  is associative and commutative. \\
 In the above algebra of functions the Leibniz's rule (\ref{formula2}) given by Corollary 2.2 is valid for functions fulfilling
 the respective assumptions concerning their behaviour at $ x_{k}=0 $:
 \begin{equation}
D^{\alpha_{k}}_{k} f*g=\beta_{k}(D^{\alpha_{k}}_{k}f)*g +(1-\beta_{k})f*D^{\alpha_{k}}_{k}g
 \end{equation}
 with $ \beta_{k} \in [0,1]$ for $ k=1,...,m $. \\
 For classical derivatives acting by assumption in directions $ j=m+1,...,n$ we obtain
 for convolution (\ref{alg}) the standard form of the Leibniz's rule:
 \begin{equation}
 \partial_{j}(f*g)=(\partial_{j}f)*g+f*\partial_{j}g
 \end{equation}
 Similarly to the one-dimensional case
 investigated in the previous section  we can
 introduce also the partial differintegral operators of positive order
 for functions fulfilling suitable conditions:
 \begin{equation}
{\cal{D}}^{\alpha_{k}}_{k}f(\vec{x}):=D^{\alpha_{k}}_{k} f(\vec{x})-f(\vec{x})
\label{oparp}
 \end{equation}
 These operators  obey the Leibniz's rule for functions multiplied according to (\ref{alg}):
 \begin{eqnarray}
&{\cal{D}}^{\alpha_{k}}_{k}(f*g)= ({\cal{D}}^{\gamma_{k}}_{k}f)*g +
(\zeta^{\gamma_{k}}_{k}f)*{\cal{D}}^{\alpha_{k}-\gamma_{k}}_{k}g& \\
&{\cal{D}}^{\alpha_{k}}_{k}( f*g)=({\cal{D}}^{\gamma_{k}}_{k}f)*\zeta^{\alpha-\gamma_{k}}_{k}g+
f*{\cal{D}}^{\alpha_{k}-\gamma_{k}}_{k}g & \\
& \partial_{j}(f*g)=(\partial_{j}f)*g+f*\partial_{j}g &
 \end{eqnarray}
where $ k=1,..,m $ and $ j=m+1,...,n $ and
the function $ f $ obeys the conditions of Lemma 2.1
for the fractional order of the derivative $\gamma_{k} $ while the second function
$ g $ respectively fulfills these conditions for $ \alpha_{k}-\gamma_{k}$. \\
The first two formulas are the symmetric forms of the Leibniz's rule for fractional derivatives and the last one is standard
Leibniz's rule for partial derivatives but taken in algebra of functions defined by multiplication (\ref{alg}). \\
Now we can apply the  properties of multiplication (\ref{alg})  and fractional differentation
in construction of the stationarity-conservation laws and conserved charges
for some partial fractional equations.
\section{Examples}
\subsection{Fractional diffusion equation in 1+1 }
Let us recall the fractional diffusion equation discussed in \cite{PD,NG,WS}:
\begin{equation}
D^{\alpha}_{t} \phi(x,t)=\lambda^{2}\partial^{2}_{x}\phi(x,t)+\phi(x,0)\frac{t^{-\alpha}}{\Gamma(1-\alpha)}
\label{Nig1}
\end{equation}
where $ t>0, \;\; x \in R $ and $ 0<\alpha<1 $ describes the process of ultraslow diffusion while
the value $ 1<\alpha<2$ is used for intermediate processes \cite{PD,NG,WS}.\\
Let us focus on the case of ultraslow diffusion.
The operator of the equation contains both types of derivatives: fractional with respect to time and standard
for the spatial dimension:
\begin{equation}
\Lambda(D^{\alpha}_{t},\partial_{x})= D^{\alpha}_{t}-\lambda^{2}\partial_{x}^{2}
\end{equation}
The product of functions for this model is defined according to (\ref{alg})
and looks as follows:
\begin{equation}
f*g(x,t)=\int_{0}^{t} f(x,t-s)g(x,s)ds
\end{equation}
Using the properties of the new multiplication and of the fractional derivative (\ref{formula2}) we
construct the operator $ \Gamma $ with components:
\begin{equation}
\Gamma_{x}=\lambda^{2} {\stackrel{\leftarrow}{\partial}}_{x}-\lambda^{2}\partial_{x}
\hspace{2cm}
\Gamma_{t}=2
\end{equation}
Then the current:
\begin{eqnarray}
& J_{x}=\phi'\Gamma_{x}*\phi=\phi'\lambda^{2} {\stackrel{\leftarrow}{\partial}}_{x}*\phi -\phi'*\lambda^{2}\partial_{x}\phi & \label{curr1}\\
& J_{t}=\phi'\Gamma_{t}*\phi=2\phi'* \phi & \label{curr2}
\end{eqnarray}
obeys the stationarity-conservation equation for $ t\geq 0 $ in the area
$ \phi(x,0)=\phi'(x,0)=0 $
\begin{equation}
\partial_{x} J_{x}+D^{\alpha}_{t} J_{t}=0
\label{cons}
\end{equation}
provided the function $ \phi $ is the solution of initial equation (\ref{Nig1})
while $ \phi' $ solves its conjugation:
\begin{equation}
\Lambda(-D^{\alpha}_{t}, -\partial_{x})\phi'(x,t)+\phi'(x,0)\frac{t^{-\alpha}}{\Gamma(1-\alpha)}=0
\end{equation}
Before passing to the proof of the stationarity-conservation law (\ref{cons})
we shall discuss the existence of solutions of  diffusion equation and of its conjugated form
with required properties around $ t=0 $.\\
Let us recall the form of general solution of the equation (\ref{Nig1}) \cite{WS}:
\begin{equation}
\phi(x,t)=\int_{-\infty}^{\infty} dy G_{\alpha}(x,y,t)\phi(y,0)
\label{solv}
\end{equation}
where $ G_{\alpha} $ is the fractional Green's function of the following form:
\begin{equation}
G_{\alpha}(x,y,t)=t^{-\alpha}\int_{0}^{\infty}dz E_{\alpha}(t^{-\alpha}z)G(x,y,z)
=\int_{0}^{\infty}dv E_{\alpha}(v)G(x,y,t^{\alpha}v)
\end{equation}
with the function $ G(x,y,z)=G( \mid x-y\mid , z) $ being the standard Green's function:
$$G( \mid x-y\mid , z)=\frac{1}{\sqrt{4 \pi z}}e^{-\frac{\mid x-y \mid^{2}}{4z}} $$
and $ E_{\alpha} $ denoting the Mittag-Leffler function \cite{MR,WS}. \\
Taking into account the asymptotic properties of the function $ G $ we conclude that the solution $ \phi $
behaves in the neighbourhood of $ t=0 $ as the power function $ t^{-\frac{\alpha}{2}} $.
The solution $ \phi' $ of the conjugated equation has a similar form
so its behaviour for $ t \rightarrow 0 $ is the same as of the considered solution $ \phi $. \\
This fact implies that at least for $ 0<\alpha<\frac{2}{3} $ the assumptions of the Lemma 2.1
are fulfilled
therefore we can use in the proof of the stationarity-conservation equation (\ref{cons})
the Leibniz's rule for fractional derivative $ D^{\alpha}_{t}$ given in formula (\ref{formula2}).\\
Let us check the conservation law explicitly applying the Leibniz's rule (\ref{formula2}) with $ \beta=\frac{1}{2} $:
\begin{equation}
\partial_{x} J_{x}+D^{\alpha}_{t} J_{t}=
\end{equation}
$$
 \partial_{x}\left (\phi'\lambda^{2} {\stackrel{\leftarrow}{\partial}}_{x}*\phi -\phi'*\lambda^{2}\partial_{x}\phi \right )
+D^{\alpha}_{t}\left (2\phi'* \phi \right)=
$$
$$ \lambda^{2} (\partial_{x}^{2}\phi')*\phi-\phi'*\lambda^{2}\partial^{2}_{x} \phi
-(-D^{\alpha}_{t}\phi')*\phi
+\phi'*D^{\alpha}_{t}\phi =$$
$$-\left [ \left (-D^{\alpha}_{t}
-\lambda^{2} \partial^{2}_{x}
\right )\phi'\right ]*\phi+\phi'* \left (D^{\alpha}_{t} -\lambda^{2} \partial^{2}_{x}\right )\phi =0$$
We have omitted the terms depending on initial values $ \phi(x,0) $ and $ \phi'(x,0) $
as we expect the rule (\ref{cons}) to be fulfilled in the area
where $ \phi(x,0)=\phi'(x,0)=0 $. \\
Having obtained the general form of
 stationary current (\ref{curr1},\ref{curr2}) we can discuss the possible symmetries
of equation (\ref{Nig1}) which can be used in construction of
different solutions of the initial diffusion problem.
The set includes spatial momentum
$P_{x}=\partial_{x} $
as this operator
commuts with the operator of diffusion equation (\ref{Nig1}).\\
The stationarity-conservation laws for currents including new solutions
are  fulfilled
for transformed solution
$ P_{x}\phi $ in the area
$ \partial _{x} \phi(x,0)=\phi'(x,0)=0 $. The simplest possible choice of initial value
for solutions of diffusion equation and of its conjugation is $ \frac{\phi(x,0)}{\phi_{0}}
=\frac{\phi'(x,0)}{\phi'_{0}}=\delta(x) $ with $ \phi_{0} $ and $ \phi'_{0} $
arbitary constants. \\
In this way we arrive at the stationary (for $ x\neq 0 $ in the sense of (\ref{cons}))
current connected with symmetry of the fractional
diffusion equation:
\begin{eqnarray}
 J^{x}_{x}=\phi'\Gamma_{x}*P_{x}\phi &
& J^{x}_{t}=\phi'\Gamma_{t}*P_{x}\phi
\end{eqnarray}
The stationarity-conservation equation (\ref{cons}) can be reformulated
using the definition of Riemann-Liouville derivative so as to obtain
the standard conservation equation namely:
\begin{equation}
\partial_{x}J'_{x}+\partial_{t}J'_{t} =0
\label{cons1}
\end{equation}
which is fulfilled for $ x\neq 0 $ and
the components of the new current look as follows:
\begin{equation}
J'_{x}=J_{x} \hspace{2cm} J'_{t}=J_{t}*\Phi_{\alpha}=\frac{1}{\Gamma(1-\alpha)}J_{t}*t^{-\alpha}
\end{equation}
Following the classical field theory the time-components of the conserved currents $ J $ and $ J' $
yield the charges:
\begin{eqnarray}
& Q=\int_{-\infty}^{\infty} dx \;\; J_{t} & \label{charge1} \\
& Q'=\int_{-\infty}^{\infty} dx \;\; J'_{t} &   \label{charge2}
\end{eqnarray}
The respective derivatives of the above charges
 are determined by
 the boundary terms for time-components of the currents and the initial conditions
 for solutions $ \phi $ and $ \phi' $:
 \begin{equation}
  D^{\alpha}_{t} Q=\lim_{x \rightarrow \infty}  \left [
\lambda^{2}(\partial_{x}\phi')*\phi -\lambda^{2} \phi'*\partial_{x}\phi\right]+   \label{term1}
\end{equation}
$$
-\lim_{x \rightarrow -\infty}  \left [
\lambda^{2}(\partial_{x}\phi')*\phi -\lambda^{2} \phi'*\partial_{x}\phi\right]+
$$
$$
+\phi'_{0}\frac{t^{-\alpha}}{\Gamma(1-\alpha)}*\phi(0,t)+\phi'(0,t)*\phi_{0}\frac{t^{-\alpha}}{\Gamma(1-\alpha)}
$$
\begin{equation}
 \frac{d}{dt} Q'= D^{\alpha}_{t} Q
  \label{term2}
  \end{equation}
From he general form of solutions (\ref{solv}) we obtain for initial condition $ \phi(x,t=0)=\phi_{0}\delta(x)$:
$$ \phi(0,t)=\phi_{0}G_{\alpha}(0,0,t) $$
and for conjugated equation:
$$ \phi'(0,t)=-\phi'_{0}G_{\alpha}(0,0,t)$$
Due to this property of the solutions and  the commutativity of the convolution
the last terms in the above formulas cancel. The first parts vanish by the asymptotic
properties of the Green's function which decays exponentially together with its spatial derivative
for large $ x $.\\
Thus the explicit expressions for charges (\ref{charge1},\ref{charge2}) produce the stationary function
$ Q$  and constant function $ Q'$ connected with the stationarity law and conservation law of the diffusion equation in 1+1 dimensions:
\begin{equation}
D^{\alpha}_{t}Q=0 \hspace{2cm} \frac{d}{dt} Q'=0
\end{equation}
\subsection{Generalized fractional diffusion}
Let us now extend the dimension of the space-like coordinates to $ d $. We shall consider the equation
known as the generalized fractional diffusion problem \cite{HS,HP}:
\begin{equation}
D^{\alpha}_{t} \phi(\vec{x},t)=C \triangle \phi(\vec{x},t)+\phi(\vec{x},0)\frac{t^{-\alpha}}{\Gamma(1-\alpha)}
\label{dif1}
\end{equation}
with $ t>0 $, $ 0<\alpha<1 $ and $ \triangle $ the Laplace operator in $ d$-dimensional Euclidean space.\\
We again check the properties in the neighbourhood of $ t=0 $ of the solutions.    \\
The  solution for arbitrary initial condition generalizes the formula (\ref{solv})
used in the previous section:
\begin{equation}
\phi(\vec{x},t)=\int_{-\infty}^{\infty} d^{d}\vec{y} G_{\alpha}(\vec{x},\vec{y},t)\phi(\vec{y},0)
\label{solv1}
\end{equation}
where $ G_{\alpha} $ is the fractional Green's function of the following form:
\begin{equation}
G_{\alpha}(\vec{x},\vec{y},t)=t^{-\alpha}\int_{0}^{\infty}dz E_{\alpha}(t^{-\alpha}z)G(\vec{x},\vec{y},z)
\end{equation}
with the function $ G(\vec{x},\vec{y},z)=G( \mid \vec{x}-\vec{y}\mid , z) $ being the standard Green's function:
$$G( \mid \vec{x}-\vec{y}\mid , z)=(4 \pi z)^{-\frac{d}{2}}e^{-\frac{\mid \vec{x}-\vec{y} \mid^{2}}{4z}} $$
and $ E_{\alpha} $ denoting the Mittag-Leffler function. \\
Taking into account the fact that this function for $ 0<\alpha<1 $
is an entire function and vanishes exponentially for large positive values of argument
we conclude that the solution $ \phi $ behaves in the neighbourhood of $ t=0 $
as the power function $ t^{-\alpha} $. Similar argument applies
to the solution of the conjugated equation $ \phi' $ given below (\ref{conjx}).\\
The product of functions given by (\ref{alg}) has in $ d+1 $-dimensional case
the following explicit form:
\begin{equation}
f*g(\vec{x},t)=\int_{0}^{t} f(\vec{x},t-s)g(\vec{x},s)ds
\end{equation}
The number of the components of the operator $ \Gamma$ and of the current $ J $ is now $ d+1 $
while the form of the space-like and time-like parts is identical to the ones
obtained for the modified Nigmatullin's diffusion equation. \\
The operator $ \Gamma $ given by:
\begin{equation}
\Gamma_{i}=C {\stackrel{\leftarrow}{\partial}}_{x_{i}}-C\partial_{x_{i}}
\hspace{2cm}
\Gamma_{t}=2
\end{equation}
can be applied in the construction of the current:
\begin{eqnarray}
& J_{i}=\phi'\Gamma_{i}*\phi=\phi'C {\stackrel{\leftarrow}{\partial}}_{x_{i}}*\phi -\phi'*C\partial_{x_{i}}\phi & \label{difc1}\\
& J_{t}=\phi'\Gamma_{t}*\phi=2 \phi'* \phi& \label{difc2}
\end{eqnarray}
where $ \phi $ solves the initial generalized diffusion equation (\ref{dif1})
and $ \phi' $ its conjugation :
\begin{equation}
\Lambda(-D^{\alpha}_{t}, -\partial_{x_{1}},...,-\partial_{x_{d}})\phi'(\vec{x},t)+\phi'(\vec{x},0)\frac{t^{-\alpha}}{\Gamma(1-\alpha)}=0
\label{conjx}
\end{equation}
The current (\ref{difc1},\ref{difc2}) obeys the stationarity-conservation equation:
\begin{equation}
\sum_{i=1}^{d}\partial_{x_{i}} J_{i}+D^{\alpha}_{t} J_{t}=0
\label{consd}
\end{equation}
for $ \vec{x} \neq \vec{0} $ provided the solution of diffusion equation (\ref{dif1})
 with the initial condition
$ \phi(\vec{x},0)=\phi_{0}\delta(\vec{x})$ is taken and for the conjugated equation the initial condition
$\phi'(\vec{x},0)=\phi'_{0}\delta(\vec{x}) $ is considered.\\
The proof of the above conservation law is analogous to the one presented in the previous section
for the $ 1+1 $ diffusion equation.
The essential feature in the proof are the asymptotic properties
of the solutions $ \phi $ and $ \phi' $ in the neighbourhood of $ t=0 $.
Similarly to the previous case they allow us to apply the Leibniz's rule (\ref{formula2})
for $\vec{x} \neq \vec{0} $ at least when $ 0<\alpha<\frac{1}{2}$.\\
The set of symmetry operators for $ d+1 $ case is much wider as it contains not only momenta:
\begin{equation}
P_{i}=\partial_{x_{i}}
\end{equation}
but also the angular momentum with respect to the space-like dimensions:
\begin{equation}
M_{ij}=x_{i}\partial_{x_{j}}-x_{j}\partial_{x_{i}}
\end{equation}
where $ i,j=1,...,d $. \\
As the symmetry operators transform solutions of (\ref{dif1}) into solutions
with the same properties around $ t=0 $
 we can use them in construction of the stationary-conserved currents:
\begin{eqnarray}
& J^{ \delta}_{i}=\phi'\Gamma_{i}*\delta\phi=\phi'C {\stackrel{\leftarrow}{\partial}}_{x_{i}}*\delta \phi
 -\phi'*C\partial_{x_{i}}\delta \phi & \label{curd1}\\
& J^{ \delta}_{t}=\phi'\Gamma_{t}*\delta\phi=2 \phi'* \delta \phi & \label{curd2}
\end{eqnarray}
where $ \delta $ denotes one of the above symmetry operators of the equation (\ref{dif1}).\\
The currents (\ref{curd1},\ref{curd2}) can be transformed into the components $ J'$
similarly as in the case of the $ 1+1 $ fractional diffusion. Taking the components:
\begin{equation}
J'^{\delta}_{i}=J^{\delta}_{i} \hspace{2cm} J'^{\delta}_{t}=J^{\delta}_{t}*\Phi_{\alpha}=
\frac{1}{\Gamma(1-\alpha)}J^{\delta}_{t}*t^{-\alpha}
\end{equation}
we obtain the conservation law for $ d+1 $ fractional diffusion process:
\begin{equation}
\sum_{i=1}^{d}\partial_{x_{i}} J'^{\delta}_{i}+\partial_{t} J'^{\delta}_{t}=0
\label{dcons}
\end{equation}
valid for $ \vec{x} \neq \vec{0} $.\\
Finally the derived set of stationary and conserved currents yields two sets of
charges indexed by the symmetry operators $ \delta \in \{P_{i}, M_{ij}\}\;\; i,j=1,...,d$:
\begin{eqnarray}
              & Q^{\delta}=\int_{\infty}^{\infty} d^{d}\vec{x}\;\; J^{\delta}_{t}&\\
              & Q'^{\delta}=\int_{\infty}^{\infty} d^{d}\vec{x}\;\; J'^{\delta}_{t} &
\end{eqnarray}
which are respectively stationary and conserved functions of time:
\begin{eqnarray}
              & D^{\alpha}_{t}Q^{\delta}=0&\\
              & \frac{d}{dt} Q'^{\delta}=0 &
\end{eqnarray}
\section{Stationarity - conservation laws for some fractional partial equations}
In previous sections we have obtained the stationarity law and conservation law for some
examples of partial fractional differential equations.  \\
In the multidimensional case of diffusion equation the
general solution allows the explicit construction of the current
which obeys the stationarity-conservation law
in the area of space where the initial conditions vanish both for
solution of diffusion equation and for its conjugation.\\
We have shown for that the stationary currents
yield stationary charges which can be converted to the conserved ones. \\
The discussed example shows that the construction of possible charges stationary or conserved
is connected with the asymptotic properties of solutions
around $ 0 $ and for $ \mid\vec{x} \mid \rightarrow \infty $. \\
In the sequel we shall discuss the general construction
of stationarity-conservation laws assuming that regular
in the sense of Lemma 2.1 solutions of the respective fractional differential equations
exist at least in certain area of space.
\subsection{Mixed fractional differential and differential partial equations}
Let us consider now the general equation which contains the fractional and differential parts of the following form:
\begin{equation}
\Lambda(D,\partial) \phi=[\tilde{\Lambda}(D)+\Lambda(\partial)]\phi=
\label{gen1}
\end{equation}
$$
\left (\sum_{k=1}^{m}\tilde{\Lambda}_{k}D^{\alpha_{k}}_{k} +\sum_{l=1}^{N} \Lambda_{\mu_{1}...\mu_{l}}
\partial^{\mu_{1}}...\partial^{\mu_{l}} +\Lambda_{0}\right ) \phi=0
$$
The introduced equation has constant coefficients
(we admit also constant matrices) and generalizes the studied examples
of fractional diffusion. We shall study the construction for the homogenous
form of the equation remembering that the addition of the initial terms
similar to the ones discussed previously restricts only the area of
application of the stationarity equation and
does not change the general construction.
 We assume that for given variables $ x_{1},...,x_{m} $
the equation includes only fractional derivatives in $ \tilde{\Lambda}(D)$ while for the remaining coordinates
$ x_{m+1},...,x_{n} $ only partial derivatives appear in the operator $ \Lambda(\partial) $.\\
To derive the stationarity-conservation law we shall use the Takahashi-Umezawa method \cite{TU}
for the differential part $ \Lambda(\partial) $ and the fractional Leibniz's rule (\ref{formula2}) for the part $ \tilde{\Lambda}(D) $
containing fractional operators. \\
As we know from the discussed examples each direction of the space yields
 the component of the current which for coordinates $ x_{1},..,x_{k} $
is given by the $ \tilde{\Gamma} $ operator of the form:
\begin{equation}
\tilde{\Gamma}_{k}=2 \tilde{\Lambda}_{k}
\end{equation}
while for the part $ j=m+1,...,n $ we obtain \cite{TU}:
\begin{equation}
\Gamma_{j}=\sum_{l=1}^{N-1} \sum_{k=1}^{l} \Lambda_{j \mu_{1}...\mu_{l}} (-{\stackrel{\leftarrow}{\partial}}^{\mu_{1}})...
(-{\stackrel{\leftarrow}{\partial}}^{\mu_{k}})\partial^{\mu_{k+1}}...\partial^{\mu_{l}}
\label{gamma1}
\end{equation}
It is the well-known fact that for an arbitrary pair of functions $ f $ and $ g $
the operator  $ \Gamma $ fulfills the equality:
\begin{equation}
\sum_{j=m+1}^{n} \partial^{j} f*\Gamma_{j} g=-f\Lambda(-\stackrel{\leftarrow}{\partial}) *g+f*\Lambda(\partial)g
\label{gamma}
\end{equation}
where the multiplication is given by the convolution (\ref{alg}) and $ \Lambda(-\stackrel{\leftarrow}{\partial})$
is the conjugated operator for $ \Lambda(\partial) $ acting on the left-hand side. \\
The above property of the $ \Gamma $ operator together
with the Leibniz's rule (\ref{formula2}) for fractional derivatives (taken with parameters $ \beta_{k}=\frac{1}{2} \;\;\;
k=1,...,m$) implies the following proposition to be valid:
\begin{prop1}
Let the function $ \phi $ be an arbitrary solution of the equation (\ref{gen1})
and let $ \phi' $ solve the conjugated equation:
\begin{equation}
\phi'\Lambda(-\stackrel{\leftarrow}{D},-\stackrel{\leftarrow}{\partial}) =
\phi'[\tilde{\Lambda}(-\stackrel{\leftarrow}{D})+\Lambda(-\stackrel{\leftarrow}{\partial})]=
\end{equation}
$$
\phi'\left (-\sum_{k=1}^{m}\tilde{\Lambda}_{k}{\stackrel{\leftarrow}{D}}^{\alpha_{k}}_{k}
+\sum_{l=1}^{N} \Lambda_{\mu_{1}...\mu_{l}}
(-{\stackrel{\leftarrow}{\partial}}^{\mu_{1}})...(-{\stackrel{\leftarrow}{\partial}}^{\mu_{l}}) +\Lambda_{0}\right ) =0
$$
Then the current given by the components:
\begin{eqnarray}
 J_{k}=\phi'*\tilde{\Lambda}_{k} \phi+\phi'\tilde{\Lambda}_{k} *\phi & & k=1,...,m\\
 J_{j}=\phi'*\Gamma_{j} \phi & & j=m+1,...,n
\end{eqnarray}
fulfills the stationarity-conservation equation:
\begin{equation}
\sum_{k=1}^{m} D^{\alpha_{k}}_{k} J_{k}+\sum_{j=m+1}^{n} \partial^{j}J_{j}=0
\label{law1}
\end{equation}
provided the solutions $ \phi $ and $ \phi' $ fulfill the conditions of Lemma 2.1 in the neighbourhood
of $ x_{k}=0 \;\;\; k=1,...,m$.
\end{prop1}
Proof:\\
We check the law (\ref{law1}) explicitly:
$$\sum_{k=1}^{m} D^{\alpha_{k}}_{k} J_{k}+\sum_{j=m+1}^{n} \partial^{j}J_{j}= $$
$$\sum_{k=1}^{m} D^{\alpha_{k}}_{k} \left (\phi'*\tilde{\Lambda}_{k} \phi+\phi'\tilde{\Lambda}_{k} *\phi\right )
+\sum_{j=m+1}^{n} \partial^{j}\left (\phi'*\Gamma_{j} \phi\right) =$$
$$\sum_{k=1}^{m} (D^{\alpha_{k}}_{k} \phi')\tilde{\Lambda}_{k}* \phi+
\sum_{k=1}^{m}\phi'*\tilde{\Lambda}_{k} D^{\alpha_{k}}_{k}\phi
 -\phi'\Lambda(-\stackrel{\leftarrow}{\partial}) *\phi+\phi'*\Lambda(\partial)\phi= $$
$$ -\phi'\Lambda(-\stackrel{\leftarrow}{D},-\stackrel{\leftarrow}{\partial}) *\phi
+ \phi'*\Lambda(D,\partial) \phi=0$$
Thus for every equation of the form (\ref{gen1}) we can produce exact form of the stationary-conserved current provided the initial equation
and its conjugation have solutions which fulfill the asymptotic conditions
 at $ x_{k}=0 \;\;\; k=1,...,m $ which allow the application of the Leibniz's rule for fractional
partial derivatives
$ D^{\alpha_{k}}_{k} $. \\
The stationarity-conservation equation (\ref{law1}) can be rewritten in the form of the standard
conservation law for modified components of the above current ($m_{k}<\alpha_{k}<m_{k}+1 \;\;\;\; k=1,...,m$):
\begin{eqnarray}
J'_{k}=(\partial^{k})^{m_{k}} (J_{k} *_{k}\Phi_{\alpha_{k}-m_{k}})& & k=1,...,m  \\
J'_{j}=J_{j}& & j=m+1,...,n
\end{eqnarray}
where the convolution $ *_{k} $
is given by the formula:
\begin{equation}
f*_{k}g(\vec{x})=
\label{convp}
\end{equation}
$$\int_{0}^{x_{k}}f(\vec{x}-s\vec{e}_{k})
g(\vec{x}+(s-x_{k})\vec{e}_{k})ds_{k}
 $$
The new current $ J' $ obeys the conservation law:
\begin{equation}
\sum_{l=1}^{n}\partial^{l}J'_{l}=0
\end{equation}
\subsection{Mixed fractional sequential
and differential partial equations}
In the previous construction we have considered the fractional part of the operator including
only the first power of the corresponding partial fractional derivatives while in the
differential part we have taken an arbitrary polynomial of partial derivatives.
Let us extend the derivation of the stationarity-conservation laws to the general case containing
both polynomial of fractional derivatives and polynomial of classical partial derivatives:
\begin{equation}
\Lambda(D,\partial) \phi=[\tilde{\Lambda}(D)+\Lambda(\partial)]\phi=
\label{seq1}
\end{equation}
$$
\left (\sum_{k=1}^{M}
\tilde{\Lambda}_{\rho_{1}...\rho_{k}}D^{\alpha_{1}}_{\rho_{1}}...D^{\alpha_{k}}_{\rho_{k}}
 +\sum_{l=1}^{N} \Lambda_{\mu_{1}...\mu_{l}}
\partial^{\mu_{1}}...\partial^{\mu_{l}} +\Lambda_{0}\right ) \phi=0
$$
The derivatives with respect to the coordinates $ x_{1},...,x_{m}$ are the
fractional $ D^{\alpha_{i}}_{\rho_{i}}$
where the upper index denotes
the fractional order and the lower one the respective
partial direction. The part depending on fractional derivatives
 has now the form of partial sequential fractional
operator generalizing the sequential operator
for one-dimensional space \cite{MR}. The coefficients $ \Lambda $ and $ \tilde{\Lambda}$
are again constant matrices or numbers.
As the derivatives with respect to different coordinates do commute both types of coefficients
are fully symmetric with respect to the permutation of the set of indices. \\
To obtain the $ \Gamma $ operator fulfilling the equation (\ref{gamma})
we again use the Takahashi-Umezawa method for the differential part $ \Lambda(\partial) $
and get the components $ \Gamma_{j} $ as given by (\ref{gamma1})  whereas for $ \tilde{\Gamma} $
we have:
\begin{equation}
\tilde{\Gamma}_{k}=2 \sum_{j=1}^{M-1}\sum_{l=1}^{j} \tilde{\Lambda}_{k\rho_{1}...\rho_{j}}
(-{\stackrel{\leftarrow}{D}}^{\alpha_{1}}_{\rho_{1}})...(-{\stackrel{\leftarrow}{D}}^{\alpha_{l}}_{\rho_{l}})
D^{\alpha_{l+1}}_{\rho_{l+1}}...D^{\alpha_{j}}_{\rho_{j}}
\end{equation}
It is easy to check the analog of the formula (\ref{gamma}) for the operator $ \tilde{\Gamma}$:
\begin{equation}
\sum_{k=1}^{m}D^{\alpha_{k}}_{k}\left ( f*\tilde{\Gamma}_{k}g\right )=
-f\tilde{\Lambda}(-\stackrel{\leftarrow}{D})*g+f*\tilde{\Lambda}(D)g
\end{equation}
for an arbitrary pair of functions $ f $ and $ g $ allowing the use of the Leibniz's rule (\ref{formula2})
together with their fractional derivatives $ D^{\alpha_{l+1}}_{\rho_{l+1}}...D^{\alpha_{j}}_{\rho_{j}} g $
and $ f(-{\stackrel{\leftarrow}{D}}^{\alpha_{1}}_{\rho_{1}})...(-{\stackrel{\leftarrow}{D}}^{\alpha_{l}}_{\rho_{l}}) $.
 \\
 All the above calculations yield as a result the following proposition which describes the explicit construction
 of the stationarity - conservation law for linear sequential fractional-differential equation (\ref{seq1}):
 \begin{prop2}
Let the function $ \phi $ be an arbitrary solution of the equation (\ref{seq1})
and let $ \phi' $ be a solution of the conjugated equation in the form:
\begin{equation}
0=\phi'\Lambda(-\stackrel{\leftarrow}{D}, -\stackrel{\leftarrow}{\partial})=
\end{equation}
$$
\phi'\left ( \sum_{k=1}^{M}
\tilde{\Lambda}_{\mu_{1}...\mu_{k}}(-{\stackrel{\leftarrow}{D}}^{\alpha_{1}}_{\mu_{1}})
...(-{\stackrel{\leftarrow}{D}}^{\alpha_{k}}_{\mu_{k}})
 +\sum_{l=1}^{N} \Lambda_{\mu_{1}...\mu_{l}}
(-{\stackrel{\leftarrow}{\partial}}^{\mu_{1}})...
(-{\stackrel{\leftarrow}{\partial}}^{\mu_{l}}) +\Lambda_{0}
 \right )
$$
Then the current with the following components:
\begin{eqnarray}
J_{k}=\phi'*\tilde{\Gamma}_{k} \phi & & k=1,...,m \label{comp} \\
J_{j}=\phi'*\Gamma_{j}\phi & & j=m+1,...,n
\end{eqnarray}
obeys the stationarity-conservation equation:
\begin{equation}
\sum_{k=1}^{m} D^{\alpha_{k}}_{k}J_{k}+\sum_{j=m+1}^{n} \partial^{j}J_{j}=0
\label{consd1}
\end{equation}
provided the solutions $ \phi $ , $ \phi' $ together with their derivatives appearing in the formulas
for components (\ref{comp})
fulfill the conditions of Lemma 2.1 in the neighbourhood
of $ x_{k}=0 \;\; k=1,..,m$
 \end{prop2}
Proof: \\
We use the properties of the solutions and of the operators $ \Gamma $ and $ \tilde{\Gamma} $
and obtain:
\begin{eqnarray}
& \sum_{j=m+1}^{n}\partial^{j}J_{j}= \sum_{j=m+1}^{n} \partial^{j} \left (\phi'*\Gamma_{j} \phi \right )=
-\phi'\Lambda(-\stackrel{\leftarrow}{\partial}) *\phi + \phi' *\Lambda(\partial)\phi & \nonumber \\
&  \sum_{k=1}^{m}D^{\alpha_{k}}_{k} J_{k}=
\sum_{k=1}^{m}D^{\alpha_{k}}_{k} \left (\phi'*\tilde{\Gamma}_{k}\phi \right )=
-\phi'\tilde{\Lambda}(-\stackrel{\leftarrow}{D})*\phi+\phi'*\tilde{\Lambda}(D)\phi
& \nonumber
\end{eqnarray}
Thus the left-hand side of the stationarity-conservation formula is of the form:
$$
\sum_{k=1}^{m}D^{\alpha_{k}}_{k} J_{k}+\sum_{j=m+1}^{n}\partial^{j}J_{j}=
$$
$$
-\phi'\left (\tilde{\Lambda}(-\stackrel{\leftarrow}{D})
+\Lambda(-\stackrel{\leftarrow}{\partial})+\Lambda_{0}\right )*\phi
+\phi'*\left (\tilde{\Lambda}(D)+\Lambda(\partial)+\Lambda_{0}\right )\phi=0
$$
and vanishes on shell.        \\
\\
We can rewrite the stationarity-conservation law to have the conservation law connected with the equation (\ref{seq1}).
To this aim we apply the definition of the Riemann-Liouville fractional derivative (\ref{dpar}).
The modified components of the current have the form similar to the one derived in the previous section
($ m_{k}<\alpha_{k}<m_{k}+1 \;\;\; k=1,...,m$):
\begin{eqnarray}
J'_{k}=(\partial^{k})^{m_{k}} \left (J_{k} *_{k}\Phi_{\alpha_{k}-m_{k}}\right)& & k=1,...,m \\
J'_{j}=J_{j}& & j=m+1,...,n
\end{eqnarray}
with the convolution $ *_{k} $ given by (\ref{convp}). \\
They obey the conservation law:
\begin{equation}
\sum_{l=1}^{n} \partial^{l} J'_{l}=0
\end{equation}
\subsection{Stationary and conserved charges for mixed fractional-differential models}
Following the results obtained for fractional diffusion
we shall apply the derived stationarity-conservation law in construction of stationary charges.\\
Two cases should be considered: when the time-derivative  in the operator of the equation
is a fractional and when it is standard partial one. \\
Let us assume that the time-derivative in equations (\ref{gen1},\ref{seq1}) is a fractional one. Integrating the time-component of the current fulfilling
the stationarity-conservation equation (\ref{law1},\ref{consd1}) we arrive at the charge:
\begin{equation}
Q=\int_{R^{n-1}}d\vec{x} \;\; J_{t}(\vec{x},t)
\end{equation}
which is a stationary function of order $ \alpha_{t}$ which also determines
the order of the fractional time-derivative:
\begin{equation}
D^{\alpha_{t}}_{t}Q=0
\end{equation}
provided the respective boundary terms vanish. For components $ J_{j} \;\; j=m+1,...,n$
it means that they vanish at the infinity in the given directions while for components
$ J_{k} \;\; k=2,..,m $ the asymptotic condition has the form:
\begin{equation}
\lim_{\mid x_{k}\mid \rightarrow \infty}(\partial^{k})^{m_{k}} (J_{k}*_{k}\Phi_{\alpha_{k}-m_{k}} ) =0
\end{equation}
where $ m_{k}<\alpha_{k}<m_{k}+1 $. \\
The second possibility is the model with standard time-derivative. Then the charge:
\begin{equation}
Q=\int_{R^{n-1}} d\vec{x} \;\; J_{t}(\vec{x},t)
\end{equation}
is a strictly stationary function of time that means it is a true constant function:
\begin{equation}
\partial^{t} Q=0
\end{equation}
when the asymptotic conditions for respective components of the currents
are fulfilled:
\begin{eqnarray}
\lim_{\mid x_{j}\mid \rightarrow \infty} J_{j} =0 & & j=m+2,...,n \\
\lim_{\mid x_{k}\mid \rightarrow \infty}(\partial^{k})^{m_{k}}( J_{k}*_{k}\Phi_{\alpha_{k}-m_{k}})=0 & &
k=1,...,m \\
\end{eqnarray}
The exact form of the symmetry algebra of the equations (\ref{gen1},\ref{seq1}) vary for different examples. Let us however
notice that it includes for all of them the momenta:
\begin{eqnarray}
P_{k}=D^{\alpha_{k}}_{k} & &k=1,...,m \label{m1}\\
P_{j}=\partial^{j} & & j=m+1,...,n    \label{m2}
\end{eqnarray}
as they commute with the operator of these equations. \\
However if we propose to use the above momenta in derivation of conserved currents and charges
we must additionally assume the regular behaviour of the $ W(D)P_{k}\phi $ and $ W(D)P_{j} \phi $
functions in the neighbourhood of 0 with respect to the $ x_{1},...,x_{m} $ coordinates ($ W(D)$
denote the polynomials of fractional derivatives appearing in the formula for
 $ \tilde{\Gamma} $ operator).\\
When this assumption is fulfilled the stationary-conserved currents
look as follows:
\begin{eqnarray}
J^{\delta}_{k}=\phi'*\tilde{\Gamma}_{k} \delta \phi & & k=1,...,m \\
J^{\delta}_{j}=\phi'*\Gamma_{j} \delta \phi & & j=m+1,..,n
\end{eqnarray}
where the operators $ \tilde{\Gamma} $ and $ \Gamma $ are given explicitly in previous sections. \\
In this case we have the family of stationary (or respectively conserved charges) depending which of
the considered two cases apply to our model.
They have the following  explicit form:
\begin{equation}
Q^{\delta}=\int_{R^{n-1}} d\vec{x} \;\; \phi'*\tilde{\Gamma}_{t} \delta \phi
\end{equation}
for the case where the time-derivative is fractional and for the standard
time derivative we have:
\begin{equation}
Q^{\delta}=\int_{R^{n-1}} d\vec{x} \;\; \phi'*\Gamma_{t} \delta \phi
\end{equation}
where $ \delta $ is one of the momentum operators given in (\ref{m1},\ref{m2}).
\section{Conclusions}
We have discussed the Leibniz's rule for the algebra of Laplace convolution
of differintegrable functions. \\
The derived procedure for construction of the nonlocal stationary currents applies
to fractional differential linear equations including Riemann-Liouville
fractional and classical derivatives
provided there exist the regular solutions of initial and conjugated equation. Similar method
is being investigated also for Weyl fractional derivatives for algebra
of functions defined by Fourier convolution.
It seemes that it can be extended to models with fractional derivatives
defined via generalized functions approach as well.\\
For the general case we have extracted the explicit form of stationary-conserved currents
assuming the regularity of solutions in the neighbourhood of $ 0 $. It was shown
that the stationary currents are connected with the conserved ones. Both types of currents produce charges:
namely stationary currents yield the stationary charges and respectively
from conserved nonclocal currents we obtain integrals of motion.

\end{document}